\documentclass[prd,amsmath,amssymb,superscriptaddress,nofootinbib,onecolumn]{revtex4}
\usepackage{amsfonts}
\usepackage{multirow}
\usepackage{makecell}
\usepackage{mathrsfs}
\usepackage{graphicx}
\usepackage{amsmath}
\usepackage{amssymb}
\usepackage{bm}
\usepackage{bbm}
\usepackage{color}
\usepackage{ulem}
\usepackage{array}
\usepackage{diagbox}
\usepackage{float}
\allowdisplaybreaks[4] %%%Allow long expressions displayed with break across the pages

\newcommand{\nn}{\nonumber}

\newcommand{\beq}{\begin{equation}}
\newcommand{\eeq}{\end{equation}}
\newcommand{\bqa}{\begin{eqnarray}}
\newcommand{\eqa}{\end{eqnarray}}

\newcommand{\bseq}{\begin{subequations}}
\newcommand{\eseq}{\end{subequations}}

\makeatletter

\begin{document}

\title{Confronting perturbative QCD with the hardest exclusive reactions: kaon electromagnetic form factors}
% Force line breaks with \\
%-------------------------
\author{Long-Bin Chen~\footnote{chenlb@gzhu.edu.cn}}
\affiliation{ School of Physics and Materials Science, Guangzhou University, Guangzhou 510006, China \vspace{0.2cm}}
%-------------------------
\author{Wen Chen~\footnote{chenwenphy@gmail.com}}
\affiliation{ Key Laboratory of Atomic and Subatomic Structure and Quantum Control (MOE),
Guangdong Basic Research Center of Excellence for Structure and Fundamental Interactions of Matter,
Institute of Quantum Matter, South China Normal University, Guangzhou 510006, China \vspace{0.2cm}}
\affiliation{Guangdong-Hong Kong Joint Laboratory of Quantum Matter, Guangdong Provincial Key Laboratory of Nuclear Science,
Southern Nuclear Science Computing Center, South China Normal University, Guangzhou 510006, China\vspace{0.2cm}}
%-------------------------
\author{Feng Feng~\footnote{f.feng@outlook.com}}
\affiliation{China University of Mining and Technology, Beijing 100083, China\vspace{0.2cm}}
\affiliation{Institute of High Energy Physics, Chinese Academy of Sciences, Beijing 100049, China\vspace{0.2cm}}
%-------------------------
\author{Yu Jia~\footnote{jiay@ihep.ac.cn}}
\affiliation{Institute of High Energy Physics, Chinese Academy of Sciences, Beijing 100049, China\vspace{0.2cm}}
\affiliation{School of Physical Sciences,
University of Chinese Academy of Sciences, Beijing 100049, China\vspace{0.2cm}}
%-------------------------
\date{\today}

%%%%%%%%%%%%%%%%%%%%%%%%%%%%%%%%%%%%%%%%%%%%%%%%%%%%%%%%%%%%%%%%%%%%%%%%%%%%%%
\begin{abstract}
Among countless channels of hard exclusive reactions, the kaon electromagnetic form factors (EMFFs)
are of special interest, which have been measured up to $Q^2 \sim 50\;{\rm GeV}^2$ in the timelike domain. 
The kaon EMFFs thereby serve an ideal platform to critically examine the
validity and effectiveness of perturbative QCD (pQCD) in accounting for hard exclusive processes.
In this work we confront the pQCD predictions that incorporate the next-to-next-to-leading-order (NNLO) perturbative corrections,
with the available kaon EMFFs data set from experimental measurements and from lattice predictions.
The inclusion of the NNLO corrections turns out to have a substantial and positive impact.  
If the profiles of the kaon light-cone distribution amplitudes (LCDAs) are taken from the recent lattice QCD prediction by 
{\tt LPC} Collaboration, the satisfactory agreement between theory and data 
can be reached for both charged and neutral kaons, in both spacelike and timelike large-$Q^2$ domains.
\end{abstract}

\maketitle

%%%%%%%%%%%%%%%%%%%%%%%%%%%%%%%%%%%%%%%%%%%%%%%%%%%%%%%%%%%%%%%%%%%%%%%%%%%%%%
\section{Introduction}
%%%%%%%%%%%%%%%%%%%%%%%%%%%%%%%%%%%%%%%%%%%%%%%%%%%%%%%%%%%%%%%%%%%%%%%%%%%%%%

The electromagnetic form factors (EMFFs) of hadrons, which probe the internal charge distributions inside a hadron,
play a fundamental role in advancing our knowledge toward Quantum Chromodynamics (QCD) and
hadron properties. To date, the most intensively studied examples reside in the light hadron sector,
exemplified by the EMFFs of the nucleons and the light pseudoscalar mesons such as pion and kaons~\cite{Gross:2022hyw}.

The kaons, comprising the charged $K^\pm$ and neutral $K^0/\overline{K}^0$,
are the next-to-lightest mesons within the $SU_f(3)$ octet pseudoscalar family.
As unique carriers of the strange quark, the kaons have historically played a vital role in the making of
the Standard Model, in particular in discovering the parity and $CP$ violation
in the weak interaction sector.
On the other hand,  the revelation of the approximate $SU(3)$ flavor symmetry in strong interaction sector
stands as a cornerstone in the eventual formulation of QCD.
Owing to the hierarchy $m_{u,d} \ll m_s \lesssim \Lambda_{\text{QCD}}$,
the conservation of $U$ and $V$ spins in static properties, such as masses, magnetic moments, and decay constants,
are satisfied with an accuracy of approximately $20\%$, in stark contrast with the nearly perfect (percent level)
precision of isospin symmetry.

Experimental investigations of the charged kaon EMFFs have been continuously pursued since 70s~\cite{Bernardini:1973pe, DM2:1988obi, CLEO:2005tiu, CMD-2:2008fsu, Seth:2012nn, BaBar:2013jqz, BaBar:2015lgl, Serednyakov:2015ooa, BESIII:2018ldc}, while the $K^{0}/\bar{K}^0$ EMFFs have also been the
subject of numerous studies since the beginning of this century~\cite{Akhmetshin:2002vj,Seth:2013eaa,BaBar:2014uwz,CMD-3:2016nhy, BESIII:2021yam,BESIII:2023zsk}.
On the theoretical front, the kaon EMFFs have also been extensively studied using various approaches~\cite{Wu:2008yr, Raha:2008ve, Raha:2010kz, Gao:2017mmp, Krutov:2016luz, Stamen:2022uqh, Ahmed:2023zkk, Xu:2023izo, Bruch:2004py, Miramontes:2022uyi}.
Governed by the soft strong interactions, the kaon EMFFs at low $Q^2$ have been explored within chiral perturbation theory~\cite{Gasser:1983yg},
which enables the extraction of the kaon charge radius.
Conversely, at large momentum transfer, the kaon EMFFs are anticipated to be accurately captured by perturbative QCD within the context of collinear factorization~\cite{Lepage:1979zb,Lepage:1979za,Lepage:1980fj,Efremov:1978rn,Efremov:1979qk,Duncan:1979ny,Duncan:1979hi} (for a comprehensive review on application of collinear factorization to the hard exclusive reactions, see \cite{Chernyak:1983ej}).

It is interesting to realize that the largest $Q^2$ in kaon EMFFs measurement has surpassed 50 ${\rm GeV}^2$~\cite{BaBar:2015lgl},
which is significantly greater than 10 ${\rm GeV}^2$ in the measured pion EMFFs~\cite{Bebek:1977pe,BaBar:2012bdw},
also considerably larger than 30 ${\rm GeV}^2$ in the measured proton EMFFs~\cite{Sill:1992qw}. We note that,
historically the applicability of pQCD at moderate momentum transfer has been subject to controversy~\cite{Isgur:1989cy,Bolz:1996sw,Radyushkin:1998rt,Braun:1999uj},
with the crux centered on whether the soft nonfactorizable mechanisms may dominate the pQCD contributions or not.
With $Q^2$ reaching about 50 ${\rm GeV}^2$, the kaon EMFFs may be viewed as the hardest exclusive reaction ever measured experimentally,
so that the pQCD should be reliably applied to such situation. Consequently the kaon EMFFs might serve an ideal arena 
to critically testify the validity and effectiveness of pQCD for hard exclusive processes.

One may wonder whether the kaon EMFFs are identical to the pion EMFFs within the projected 20\% accuracy.
Intriguingly, the effects of $SU(3)$ flavor symmetry breaking in these observables turn to be more pronounced than
those in static properties.
Actually there has existed a long-standing puzzle concerning the kaon EMFFs:
although the energy dependence of higher-energy data aligns with the asymptotic form predicted by pQCD,
their magnitudes exceed the predicted asymptotic value by a factor of approximately four~\cite{BaBar:2013jqz,BaBar:2015lgl}.

Recently the next-to-next-to-leading-order (NNLO) perturbative corrections to pion EMFFs at large momentum transfer
have become available in the context of collinear factorization~\cite{Chen:2023byr}.
It was found that the effect of the NNLO correction is positive and substantial. In comparison with the data,
some strong constraint has been placed on the profile of the leading-twist
light-cone distribution amplitude (LCDA) of pion, especially on the second Gegenbauber moment of the pion LCDA.

Due to the mass hierarchy between $u$, $d$ and $s$ quark, one anticipates that the profiles of the kaon LCDAs
significantly deviate from those of the pions, which are symmetric under the exchange $x\leftrightarrow 1-x$.
Thereby the kaon EMFFs provide a prime opportunity to advance our understanding of the intricate interplay between quark masses and
$SU_f(3)$ breaking and the underlying QCD dynamics.  It is the very goal of this work to extend the preceding analysis of pion
EMFFs~\cite{Chen:2023byr} to the kaon EMFFs, by incorporating the NNLO QCD corrections.

The rest of the paper is distributed as follows.
%----------------------------------------------------
In Sec.~\ref{Two:loop:expression:kaon:EMFFs}, we sketch the pQCD formula for kaon EMFFs
in collinear factorization. With the virtue of turning a convolution task into an algebraic one,
we present the master formula for pQCD predictions to kaon EMFFs through NNLO accuracy.
%----------------------------------------------------
In Sec.~\ref{phenomenology}, we perform a comprehensive comparison between the state-of-the-art pQCD predictions and
the available large-$Q^2$ data and recent lattice predictions for the EMFFs of the charged and neutral kaons, 
in both spacelike and timelike domains.
%----------------------------------------------------
Finally we summarize in Sec.~\ref{summary}.
%----------------------------------------------------
In Appendix~\ref{appendix:A}, we present the analytic expressions of the lowest-lying 
short-distance coefficient matrix elements through NNLO accuracy, which appear
in the master formula given in Sec.~\ref{Two:loop:expression:kaon:EMFFs}.
%----------------------------------------------------

%%%%%%%%%%%%%%%%%%%%%%%%%%%%%%%%%%%%%%%%%%%%%%%%%%%%%%%%%%%%%%%%%%%%%%%%%%%%%%
\section{Two-loop pQCD corrections to kaon EMFFs}
\label{Two:loop:expression:kaon:EMFFs}
%%%%%%%%%%%%%%%%%%%%%%%%%%%%%%%%%%%%%%%%%%%%%%%%%%%%%%%%%%%%%%%%%%%%%%%%%%%%%%

To be specific, let us start with the EMFF of the $K^+$ meson. It is defined via
%----------------------------------------------------
\beq
%----------------------------------------------------
\langle K^+(P')\vert  J_{\rm em}^\mu  \vert K^+(P) \rangle = F_K(Q^2) (P^\mu + P'^\mu),
%----------------------------------------------------
\eeq
%----------------------------------------------------
where $J^\mu_{\rm em} = \sum_f e_f \bar{\psi}_f \gamma^\mu \psi_f$ signifies the electromagnetic current,
and $Q^2 \equiv -(P' - P)^2$.

At large momentum transfer, the collinear factorization demands that, at the lowest order in $1/Q$,
the $F_K(Q^2)$ can be expressed in the following convolution integral:
%----------------------------
\beq
%----------------------------
F_K(Q^2) = \int\!\!\! \int dx\,dy\,\Phi_K^*(x,\mu_F) \, T \left( x,y,\frac{\mu_R^2}{Q^2},\frac{\mu_F^2}{Q^2} \right)
\,\Phi_K(y,\mu_F)+\cdots,
%----------------------------
\label{EMFF:collinear:factorization}
%----------------------------
\eeq
%----------------------------
where $\mu_R$ and $\mu_F$ represent the renormalization and factorization scales, respectively.
$T(x,y)$ signifies the perturbatively calculable hard-scattering kernel, and $\Phi_K(x,\mu_F)$
represents the nonperturbative yet universal leading-twist kaon LCDA, $i.e.$,
the probability amplitude of finding the valence $u$ and $\bar{s}$ quarks inside $K^+$
carrying the momentum fractions $x$ and $\bar{x}\equiv 1-x$, respectively.

The leading-twist kaon LCDA in \eqref{EMFF:collinear:factorization} assumes the following operator definition:
%----------------------------
\bqa
%----------------------------
\Phi_K(x,\mu_F) && = \int \frac{d z^-}{2 \pi i} e^{i z^- x P^+}\left\langle 0 \left| \bar{s}(0) \gamma^+ \gamma_5  {\cal W}(0, z^-) u(z^-)\right| K^+(P) \right\rangle,
%----------------------------
\label{pi:LCDA:operator:def}
%----------------------------
\eqa
%----------------------------
with ${\cal W}$ signifies the light-like gauge link to ensure gauge invariance.
$\Phi_K(x,\mu_F)$ obeys the celebrated Efremov-Radyushkin-Brodsky-Lepage (ERBL) evolution equation~\cite{Lepage:1980fj, Efremov:1979qk}:
%----------------------------
\begin{equation}
%----------------------------
\frac{d\Phi_K(x,\mu_F)}{d\ln\mu^2_F} = \int_0^1\!\! dy V(x,y) \, \Phi_K(y,\mu_F).
%----------------------------
\label{ERBL:evolution:eq}
%----------------------------
\end{equation}
%----------------------------

The hard-scattering kernel can be expanded in the power series of $\alpha_s$:
%----------------------------
\beq
%----------------------------
T(x,y,\mu_R^2/Q^2, \mu_F^2/Q^2) = \frac{16C_F\pi\alpha_s}{ Q^2} \left\{ T^{(0)} + {\alpha_s\over \pi} T^{(1)} +
\left({\alpha_s\over \pi}\right)^2 T^{(2)} + \cdots \right\}.
%----------------------------
\eeq
%----------------------------

The hard-scattering kernel can be determined via standard perturbative matching strategy.
Luckily, there requires no any extra work except one directly transplants the $T^{(i)}$ ($i=0,1,2$) 
for the $\pi^+$ EMFF~\cite{Chen:2023byr} by making the substitution $e_d\to e_s$. 
For technical details, we refer the interested readers to our preceding paper~\cite{Chen:2023byr}.

The leading-twist kaon LCDA is customarily expanded in the basis of the Gegenbauer polynomials:
%----------------------------
\begin{subequations}
%----------------------------
\bqa
%----------------------------
&& \Phi_K(x,\mu_F) = \frac{f_K}{2\sqrt{2N_c}} {\sum_{n = 0}} a_n(\mu_F) \psi_n(x),
%----------------------------
\label{kaon:DA:expanded:in:polynomial:basis}
%----------------------------
\\
%----------------------------
&& \psi_n(x) = 6x\bar{x} \, C_n^{3/2}(2x-1),
%---------------------
\eqa
%---------------------
\label{kaon:DA:expanded:in:Gegenbauer}
%---------------------
\end{subequations}
%---------------------
%----------------------------
where $N_c=3$ is the number of colors, $f_K = 0.160$ GeV signifies the kaon decay constant~\cite{ParticleDataGroup:2022pth}.
All the nonperturbative dynamics is encoded in the Gegenbauer moments.
At a given factorization scale, the running Gegenbauer moment 
$a_n(\mu_F)$ is known to three-loop accuracy~\cite{Strohmaier:thesis}.

Since the V-spin symmetry has been badly broken, the sum in \eqref{kaon:DA:expanded:in:polynomial:basis}
entails all integers.  This feature should be contrasted with that for the pion LCDA,
where the sum is only extended over even integers due to nearly perfect isospin symmetry.

To facilitate phenomenological analysis, it is advantageous to turn the convolution problem into an algebraic one~\cite{Chen:2023byr}.
Substituting \eqref{kaon:DA:expanded:in:Gegenbauer} into \eqref{EMFF:collinear:factorization}, conducting two-fold integration,
one can reexpress the kaon EMFF as
%----------------------------
\beq
%----------------------------
 Q^2 F_K(Q^2) = \frac{2C_F \pi^2 f_K^2}{3}  \sum_{k=0} \left(\alpha_s\over \pi\right)^{k+1} \, {\sum_{m,n}}
 [e_u-(-1)^{m+n}e_s]\, {\cal T}^{(k)}_{mn}(\mu/Q)\,a_m(\mu) a_n(\mu).
%----------------------------
\label{Master:formula:kaon:EM:FF}
%----------------------------
\eeq
%----------------------------
The short-distance matrix element ${\cal T}^{(k)}_{mn}$ is defined by
%----------------------------
\beq
%----------------------------
    {\cal T}^{(k)}_{mn}(\mu/Q) = {1\over e_u-(-1)^{m+n}e_s}
    \int\!\!\! \int dx\,dy\, \psi_m(x)  \, T^{(k)}\left(x,y,\mu^2/Q^2\right) \, \psi_n(y),
%----------------------------
\label{T:k:mn}
 %----------------------------
\eeq
%----------------------------
where we have fixed the number of light quarks to three, and for simplicity set $\mu_R=\mu_F=\mu$ in \eqref{Master:formula:kaon:EM:FF}.

%----------------------------
\begin{table}[tbh]
%----------------------------
%\centering\setlength{\tabcolsep}{1mm}{}
\begin{tabular}{c|cc|ccc}
\hline\hline
$(m,n)$ & $c_1$ & $c_2$ & $d_1$ & $d_2$ & $d_3$ \\
\hline\hline
(0,0) & 20.25 & 59.25   & 45.5625 & 302.936 & 514.600\\
(0,1) & 28.25 & 88.7893 & 76.1181 & 540.224 & 1018.02 \\
(0,2) & 32.75 & 112.473 & 96.4306 & 735.637 & 1498.75  \\
(0,3) & 35.95 & 131.413 & 112.244 & 901.650 & 1933.00 \\
(0,4) & 38.45 & 147.638 & 125.390 & 1049.88 & 2346.49 \\
(1,1) & 36.25 & 130.750 & 113.785 & 908.006 & 1925.92  \\
(1,2) & 40.75 & 160.199 & 138.097 & 1184.52 & 2697.02 \\
(1,3) & 43.95 & 183.855 & 156.755 & 1417.99 & 3398.57  \\
(1,4) & 46.45 & 203.376 & 172.123 & 1619.63 & 4034.36  \\
(2,2) & 45.25 & 192.871 & 164.660 & 1513.95 & 3690.60  \\
(2,3) & 48.45 & 218.864 & 184.918 & 1788.70 & 4575.65  \\
(2,4) & 50.95 & 240.181 & 201.536 & 2024.03 & 5371.58  \\
(3,3) & 51.65 & 246.688 & 206.313 & 2096.55 & 5620.90  \\
(3,4) & 54.15 & 269.327 & 223.820 & 2358.13 & 6548.85  \\
(4,4) & 56.65 & 292.970 & 242.021 & 2640.85 & 7589.17  \\
%----------------------------
\hline\hline
%----------------------------
\end{tabular}
%----------------------------
\caption{The numerical values of ${\cal T}^{(1,2)}_{mn}(\mu/Q)$ for $0\le m,n\le 4$.
We have parameterized  ${\cal T}^{(1)}_{mn}(\mu/Q) = c_1 L_\mu + c_2$ and
${\cal T}^{(2)}_{mn}(\mu/Q) = d_1 L_\mu^2 + d_2 L_\mu+ d_3$, with $L_\mu \equiv \ln Q^2/\mu^2$.  }
%----------------------------
\label{tab:Tmn-numeric-values}
%----------------------------
\end{table}
%----------------------------

The two-fold convolution integrals in \eqref{T:k:mn} can be worked out in closed form at various perturbative orders.
At tree level, one simply has ${\cal T}^{(0)}_{mn} =9$ for all $m,n$.
The analytic expressions of some lowest-lying short-distance matrix elements
${\cal T}^{(1,2)}_{mn}$ have been presented In Appendix~\ref{appendix:A}.
For reader's convenience, we enumerate in Table~\ref{tab:Tmn-numeric-values} the numerical values of
${\cal T}^{(1,2)}_{mn}$ for $0\le m,n\le 4$, which appears to suffice for most phenomenological applications at current stage.
It is straightforward to adapt the master formula for the charged kaon EMFF from spacelike domain to the timelike one,
provided that one makes the replacement $L_\mu \to  L_\mu+ i\pi$ in Table~\ref{tab:Tmn-numeric-values},
with $Q^2$ now indicating the squared invariant mass of the $K^+K^-$ pair.

In the exact $U$-spin limit, the neutral kaon EMFF vanishes due to $e_d=e_s$.
The magnitudes of the neutral kaon EMFFs thereby signal the $SU(3)$ flavor symmetry breaking effect.
For the EMFF of neutral $K$ mesons, one simply makes the replacement $e_u \to e_d$ in \eqref{Master:formula:kaon:EM:FF}.
In this case, the short-distance matrix element becomes nonvanishing only when $m+n$ is an odd number.
Consequently, the leading term of ${\cal T}^{(k)}_{mn}$ starts with $(m,n)$ equal to $(0,1)$ rather than $(0,0)$.

Although it is the $K^0/\bar{K}^0$ that are produced via strong force, it is the
approximate $CP$ eigenstates $K_S/K_L$
($K_S={1\over \sqrt{2}}(\vert K^0\rangle+\vert \overline{K}^0\rangle)$, $K_L={1\over \sqrt{2}}(\vert K^0\rangle-\vert \overline{K}^0\rangle)$)
which are reconstructed via weak decay.
By charge conjugation symmetry, the following different kinds of kaon EMFFs are related to each other:
%----------------------------
\beq
%----------------------------
F_{K^+}(Q^2)=-F_{K^-}(Q^2),\quad F_{K^0}(Q^2)=-F_{\bar{K}^0}(Q^2),
%----------------------------
\quad
%----------------------------
|F_{K_L}(Q^2)|=|F_{K_S}(Q^2)|=|F_{K^0}(Q^2)|.
%----------------------------
\eeq
%----------------------------
These relations hold both in spacelike and timelike domains.

%%%%%%%%%%%%%%%%%%%%%%%%%%%%%%%%%%%%%%%%%%%%%%%%%%%%%%%%%%%%%%%%%%%%%%%%%%%%%%
\section{Confronting the state-of-the-art pQCD predictions with data}
\label{phenomenology}
%%%%%%%%%%%%%%%%%%%%%%%%%%%%%%%%%%%%%%%%%%%%%%%%%%%%%%%%%%%%%%%%%%%%%%%%%%%%%%

As the key nonperturative input, our knowledge of the leading-twist kaon LCDA is still far from complete.
Early phenomenological analysis simply uses the asymptotic form and CZ parametrization~\cite{Chernyak:1983ej}.
In recent years the kaon LCDA have been investigated from different theoretical approaches, such as light-cone sum rule~\cite{Khodjamirian:2004ga,Ball:2006wn},
instanton vacuum model~\cite{Nam:2006mb,Nam:2006sx,Nam:2006au}, holographic QCD~\cite{Chang:2016ouf},
light-front quantization~\cite{Lan:2019rba}, and recently from the lattice simulations~\cite{Braun:2006dg,RQCD:2019osh,LatticeParton:2022zqc}.
The $SU_f(3)$ symmetry breaking in kaon LCDA has also been studied within the QCD sum rule approach~\cite{Ball:2003sc,Chetyrkin:2007vm}.
The kaon's Gegenbauer moments predicted from various approaches are scattered in a wide range.

%----------------------------
\begin{table}[h]
%----------------------------
% \centering\setlength{\tabcolsep}{1mm}{}
\begin{tabular}{c|c|c|c|c}
\hline\hline
$a_i$(2\;GeV) & $a_1$ & $a_2$ & $a_3$ & $a_4$  \\
\hline
{\tt RQCD}~\cite{RQCD:2019osh} & $-0.0525^{+31}_{-33}$ & $0.106^{+15}_{-16}$ & - & -  \\
{\tt LPC}~\cite{LatticeParton:2022zqc} & $-0.108\pm0.014\pm0.051$ & $0.170\pm0.014\pm0.044$ & $-0.043\pm0.006\pm0.022$ & $0.073\pm0.008\pm0.021$ \\
\hline\hline
\end{tabular}
\caption{The values of $a_i$ ($i=1,\cdots,4$) at 2 GeV evaluated by {\tt RQCD} and {\tt LPC}.}
\label{tab:GGV}
\end{table}

In the phenomenological study, we take the recent lattice results by \texttt{RQCD}~\cite{RQCD:2019osh} and \texttt{LPC}~\cite{LatticeParton:2022zqc}
Collaborations
as the inputs for the Gegenbauer moments of the kaon LCDAs in $\overline{\rm MS}$ scheme, whose values are tabulated in
Table~\ref{tab:GGV}~\footnote{It appears that the convention in defining the kaon LCDA differs between \texttt{RQCD}~\cite{RQCD:2019osh} and \texttt{LPC}~\cite{LatticeParton:2022zqc}.
We have inserted an additional minus sign to $a_1$ in {\tt RQCD} prediction to be in conformity with the {\tt LPC} prediction.}.
These two groups of lattice studies are based on quite different algorithms.
\texttt{RQCD} has presented precise predictions of the first and second Gegenbauer moment of the kaon LCDA.
\texttt{LPC} has presented the profile of the kaon LCDA, from which various Gegenbauer moments can
be extracted, and relatively larger $SU_f(3)$-symmetry breaking effects have been observed.

We use three-loop evolution equation~\cite{Braun:2017cih,Strohmaier:thesis} to evolve each $a_n$ evaluated initially at 2 GeV by lattice simulation to any
desired scale $\mu$. We retain only those $a_n$ with $n$ up to $4$ in phenomenological exploration.
The package {\tt FAPT}~\cite{Bakulev:2012sm} is invoked to evaluate the running QCD coupling constant
to three-loop accuracy.

%----------------------------
\begin{figure}[bht]
\centering
\includegraphics[width=0.45\textwidth]{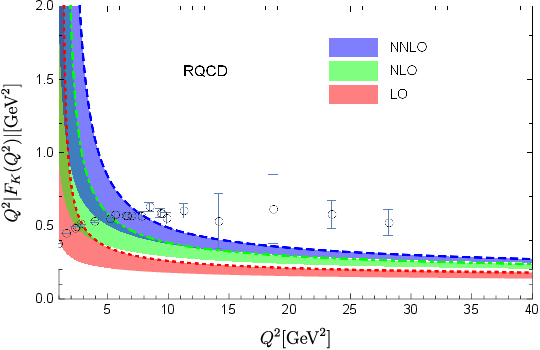}
\includegraphics[width=0.45\textwidth]{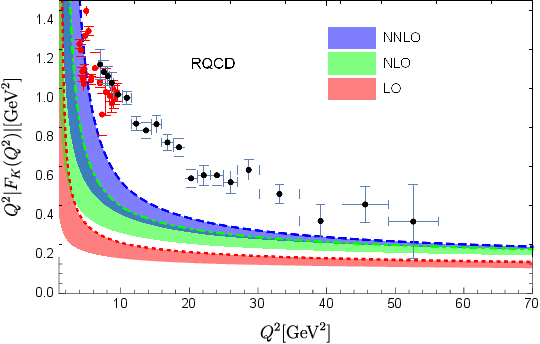}
\caption{pQCD predictions vs. experimental data for $Q^2 F_{K^{\pm}}(Q^2)$ in the space-like (left) and timelike (right) domains.
We take the central values of $a_{1,2}$ provided by \texttt{RQCD} as input in making predictions.
The red, green and blue bands correspond to the LO, NLO and NNLO predictions with the uncertainty estimated by sliding $\mu$ from $Q/2$ to $Q$.
Due to lack of experimental measurement in the spacelike domain, we take the lattice data points from the very recent lattice study~\cite{Ding:2024lfj} as substitution.
In the timelike domain, we take the data points from {\tt BaBar}~\cite{BaBar:2015lgl} (black) and {\tt BESIII}~\cite{BESIII:2018ldc}(red) experiments.
}
\label{plot:theory:vs:data:RQCD}
\end{figure}
%----------------------------

%----------------------------
\begin{figure}[bht]
\centering
\includegraphics[width=0.44\textwidth]{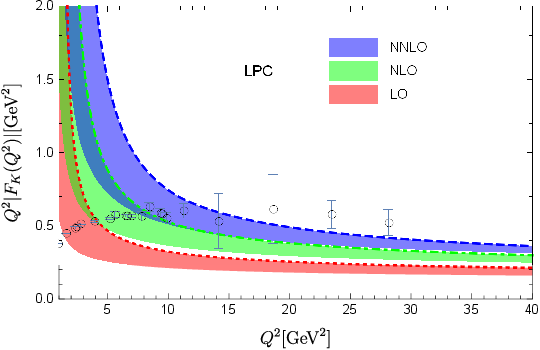}
\includegraphics[width=0.45\textwidth]{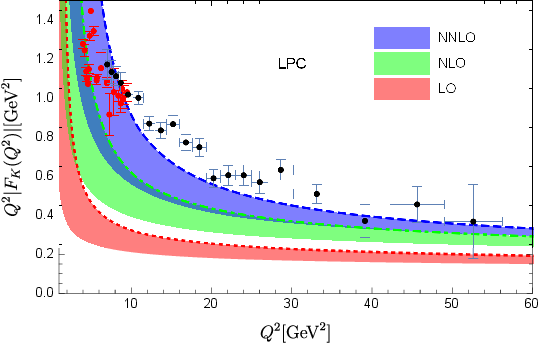}
\caption{Same as Fig.~\ref{plot:theory:vs:data:RQCD}, except the predictions are made by taking the central values of $a_i$ from  \texttt{LPC}
as inputs. }
\label{plot:theory:vs:data:LPC}
\end{figure}
%----------------------------

In Fig.~\ref{plot:theory:vs:data:RQCD} and Fig.~\ref{plot:theory:vs:data:LPC} we confront the state-of-the-art pQCD predictions
with the available data for charged kaon EMFFs, with the predictions made by taking the central values of $a_i$ given by \texttt{RQCD} and \texttt{LPC}, respectively, as
nonperturbative input parameters.
Since the experimental measurements of kaon EMFFs at spacelike domain are rather sparse,
we invoke the very recent accurate lattice determination of
the spacelike kaon EMFFs up to quite large $Q^2$~\cite{Ding:2024lfj} as an experimental substitute.
The charged kaon EMFFs in the timelike domain have recently been measured quite precisely by
{\tt BaBar}~\cite{BaBar:2015lgl} and {\tt BESIII}~\cite{BESIII:2018ldc}.
It is impressive that the covered $Q^2$ range in the former experiment has surpassed 50 ${\rm GeV}^2$,
which implies that $e^+e^-\to K^+K^-$ can be viewed as the hardest hard exclusive reaction,
an ideal platform to test the validity of the pQCD and collinear factorization.

%----------------------------
\begin{figure}[bht]
\centering
\includegraphics[width=0.4\textwidth]{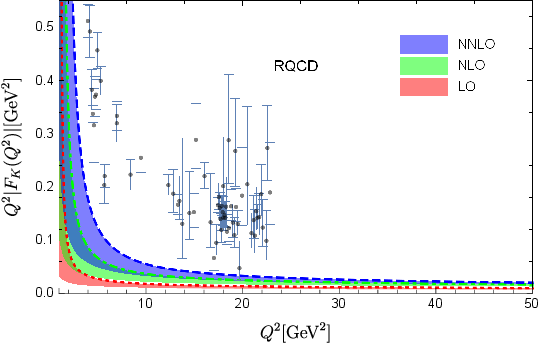}
\includegraphics[width=0.4\textwidth]{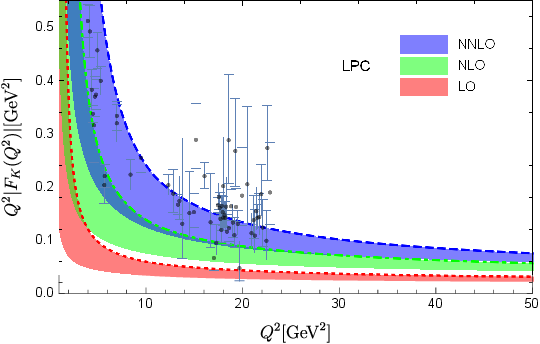}
\caption{pQCD predictions vs. {\tt BESIII} measurement of $Q^2 F_{K_S K_L}(Q^2)$ in the timelike domain.
The predictions are made by taking the central values of the Gegenbauer moments of the kaon LCDA from {\tt RQCD} (left) an {\tt LPC} (right) as inputs, respectively.
The red, green and blue bands correspond to the LO, NLO and NNLO predictions, with the uncertainty estimated by sliding $\mu$ from $Q/2$ to $Q$.}
\label{plot:theory:vs:data:LPC:K0}
\end{figure}

In Fig.~\ref{plot:theory:vs:data:LPC:K0}, we also compare the pQCD predictions of the neutral kaon EMFF in timelike domain
at various perturbative accuracy with the very recent {\tt BESIII} data~\cite{BESIII:2021yam,BESIII:2023zsk}.
Clearly the measured $F_{K_S K_L}(Q^2)$ is nonvanishing in a wide range of $Q^2$, thereby 
indicating a notable $SU_f(3)$ flavor symmetry breaking effect.

From Fig.~\ref{plot:theory:vs:data:RQCD} to  Fig.~\ref{plot:theory:vs:data:LPC:K0},
we observe a unified pattern that the NNLO perturbative corrections are always positive and substantial for both charged and neutron kaons.
Under all the circumstances, including the two-loop QCD corrections further increase the NLO predictions by a factor of two,
thereby bearing a significant phenomenological impact.
We also observe that, the NNLO predictions based on the {\tt RQCD} input appear to be significantly lower than the
lattice predictions in spacelike domain~\cite{Ding:2024lfj} and the measured timelike form factors in $e^+e^-$ collision experiments
at large $Q^2$ range, for both charged and neutral kaons. On the contrary, the NNLO predictions based on the {\tt LPC} input can satisfactorily account for
the lattice prediction and measured values for kaon EMFFs at large $Q^2$.  
The reason may be largely attributed to the fact that the value of $a_1$ given by {\tt LPC} is about 
twice greater than that by {\tt RQCD}~\footnote{It is worth reminding that, 
in the pion case, it is the NNLO prediction based on the {\tt RQCD} input that gives a better account for the measured EMFFs than that based on
the {\tt LPC} input~\cite{Chen:2023byr}.}.

To strengthen our understanding, we further confront the NNLO predictions with the
measurement of the timelike neutral kaon EMFF at $Q^2=17.4\;\text{GeV}^2$ by {\tt CLEO-\!c\!}~\cite{Seth:2013eaa}:
%---------------------------
\begin{subequations}
%---------------------------
\bqa
%---------------------------
F_{K_S K_L}(17.4\;\text{GeV}^2)|_{\tt CLEO\!-\!c} &=& 5.3\times 10^{-3}, \qquad \left[(3.1-7.9)\times 10^{-3}\;{\rm at}\;90\%\;{\rm C.L.}\right]
%---------------------------
\\
%------------------------------------
F_{K_S K_L}(17.4\;\text{GeV}^2)|^\texttt{RQCD}_{\text{NNLO}}&=& (1.4-2.1)\times 10^{-3},
%---------------------------
\\
%---------------------------
F_{K_S K_L}(17.4\;\text{GeV}^2)|^\texttt{LPC}_{\text{NNLO}}&=& (5.2-8.5)\times 10^{-3},
%---------------------------
\eqa
%---------------------------
\end{subequations}
%---------------------------
where the theoretical uncertainties are estimated by varying the renormalization and factorization scales in the range $Q/2<\mu<Q$.
Again the NNLO predictions based on the {\tt LPC} input agrees with the {\tt CLEO\!-\!c} measurement to a decent degree.

{\tt CLEO\!-\!c} also presents the ratio of the timelike EMFFs of the neutral kaon to that of the charged kaon at $Q^2=17.4\;\text{GeV}^2$~\cite{Seth:2013eaa},
which gauges the degree of the $SU(3)_f$ symmetry breaking. This measurement can be contrasted with
the NNLO predictions based on the {\tt RQCD} and {\tt LPC} inputs, with uncertainties estimated by
varying $\mu$ in the range $Q/2<\mu<Q$:
%---------------------------
\begin{subequations}
%----------------------------
\bqa
%----------------------------
{F_{K_S K_L}(17.4\;\text{GeV}^2) \over
F_{K^+ K^-}(17.4\;\text{GeV}^2)} \Big |_{\tt CLEO\!-\!c}
&=& 0.12, \qquad \left[0.07- 0.19 \;{\rm at}\;90\%\;{\rm C.L.}\right]
%----------------------------
\\
%----------------------------
\frac{F_{K_S K_L}(17.4\text{GeV}^2)}{F_{K^+ K^-}(17.4\text{GeV}^2)} \Big |^{\tt RQCD}_{\text{NNLO}}&=& 0.076-0.087,
%----------------------------
\\
%----------------------------
\frac{F_{K_S K_L}(17.4\text{GeV}^2)}{F_{K^+ K^-}(17.4\text{GeV}^2)}
\Big |^{\tt LPC}_{\text{NNLO}} &=& 0.21- 0.24.
%----------------------------
\eqa
%----------------------------
\label{ratio:CLEO-c:17.4GeV2}
%----------------------------
\end{subequations}
%---------------------------

The same ratio of the timelike EMFFs of the neutral to the charged kaons has also been measured at {\tt BESIII}
in the interval $12\;\text{GeV}^2 < Q^2 < 25\;\text{GeV}^2$~\cite{BESIII:2023zsk}:
%----------------------------
\beq
%----------------------------
{F_{K_S K_L}(12 \;\text{GeV}^2 < Q^2 < 25\;\text{GeV}^2)
\over F_{K^+ K^-}(12\;\text{GeV}^2 < Q^2 < 25\;\text{GeV}^2)
} \Big|_{\tt BESIII} = 0.21\pm 0.01.
%----------------------------
\label{bes3:measurement:ratio}
%----------------------------
\eeq
%----------------------------

%----------------------------
\begin{figure}[h]
\centering
\includegraphics[width=0.5\textwidth]{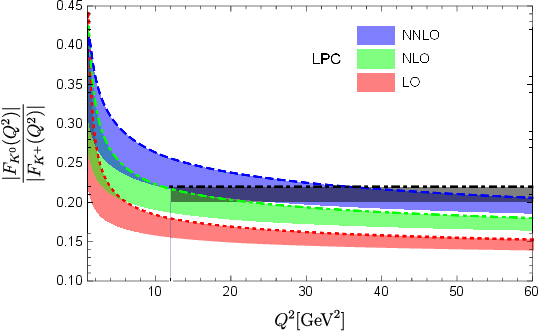}
\caption{Ratio of the EMFFs of neutral to charged kaons in the timelike domain as the function of $Q^2$.
For definiteness, we have chosen the various Gegenbauer moments of kaon LCDA with the central values of {\tt LPC} predictions.
The red, green and blue bands correspond to the LO, NLO and NNLO prediction,  with the errors estimated by sliding $\mu$ from $Q/2$ to $Q$.
The black rectangles represent the ratio measured by {\tt BESIII} measurement, \eqref{bes3:measurement:ratio}. However in the theoretical
calculation, we extend the upper limit of $Q^2$ from $25\; \text{GeV}^2$ to $60\;\text{GeV}^2$.}
\label{plot:theory:vs:data:LPC:ratio}
\end{figure}
%----------------------------

In Fig.~\ref{plot:theory:vs:data:LPC:ratio} we also juxtapose the ratio predicted by pQCD using the {\tt LPC} input and measured by 
{\tt BESIII}, \eqref{bes3:measurement:ratio}. It is evident from \eqref{ratio:CLEO-c:17.4GeV2} and Fig.~\ref{plot:theory:vs:data:LPC:ratio} that,
the pQCD predictions at NNLO accuracy, when armed with the Gegenbauber moments provided by {\tt LPC}, 
gives a satisfactory account of the {\tt CLEO\!-\!c} and {\tt BESIII} data.

%%%%%%%%%%%%%%%%%%%%%%%%%%%%%%%%%%%%%%%%%%%%%%%%%%%%%%%%%%%%%%%%%%%%%%%%%%%%%%
\section{Summary}
\label{summary}
%%%%%%%%%%%%%%%%%%%%%%%%%%%%%%%%%%%%%%%%%%%%%%%%%%%%%%%%%%%%%%%%%%%%%%%%%%%%%%

The timelike kaon EMFFs, {\it viz.}, $e^+e^- \to\gamma^*\to K\overline{K}$,
are arguably the hardest exclusive reactions ever measured experimentally, with the momentum transfer exceeding $50\;{\rm GeV}^2$. 
Therefore, the kaon EMFFs may be regarded as an ideal testing ground for the validity of pQCD when applied to hard exclusive processes.
Moreover, the magnitude of the neutral kaon's EMFF also affords a sensitive probe to the $SU(3)$ flavor
symmetry breaking effect.

In this work we conduct a comprehensive comparison between the state-of-the-art pQCD predictions in collinear factorization
and the available kaon EMFFs data sets with large momentum transfer.
For the lacking of kaon EMFF measurement in the spacelike domain,  we compare our pQCD predictions with the recent
lattice determinations of the kaon EMFF at large $Q^2$. 
In the timelike domain, we confront our predictions with the {\tt BaBar}, {\tt BESIII} and {\tt CLEO\!-\!c} data.
The key finding is that, the inclusion of the NNLO perturbative corrections has a substantial and positive impact on the theoretical predictions.
We also observe that, if the values of Gegenbauber moments of the kaon LCDAs are taken from the lattice results given by {\tt LPC},
the satisfactory agreement between theory and data (including lattice predictions to spacelike kaon EMFFs) 
can be reached for both charged and neutral kaons, in both spacelike and timelike domains with large momentum transfer.
Obviously it is highly desirable if the kaon LCDAs can be ascertained to a higher precision.

%%%%%%%%%%%%%%%%%%%%%%%%%%%%%%%%%%%%%%%%%%%%%%%%%%%%%%%%%%%%%%%%%%%%%%%%%%%%%%
\begin{acknowledgments}
%-------------------------
We thank Heng-Tong Ding and Qi Shi for providing us with the lattice predictions for the spacelike kaon EMFF in \cite{Ding:2024lfj}.
We are also indebted to Jun Hua for clarifying the sign discrepancy in $a_1$ between {\tt RQCD} and {\tt LPC} predictions.
%-------------------------
The work of L.-B.~C. is supported by the NNSFC under Grant No. 12175048.
%-------------------------
The work of W.~C. is supported by Guangdong Major Project of Basic and Applied Basic Research~(No. 2020B0301030008).
%-------------------------
The work of F.~F. is supported by the NNSFC under Grant No. 12275353.
%-------------------------
The work of Y.~J. is supported in part by the NNSFC under Grant No.~11925506.
%-------------------------
\end{acknowledgments}
%-------------------------

%%%%%%%%%%%%%%%%%%%%%%%%%%%%%%%%%%%%%%%%%%%%%%%%%%%%%%%%%%%%%%%%%%%%%%%%%%%%%%

\appendix
\section{The analytical expressions of the lowest-lying ${\cal T}^{(k)}_{mn}$}
\label{appendix:A}

The two-fold convolution integrals in (\ref{T:k:mn}) can be worked out in the closed form. At one-loop level, the
lowest-lying short-distance coefficient matrix elements read
%----------------------------
\begin{subequations}
%----------------------------
\bqa
%----------------------------
{\cal T}^{(1)}_{01} &=& \frac{113}{4}  L_\mu-\frac{18}{5}\zeta_3+\frac{5587}{60},
%----------------------------
\\
%----------------------------
{\cal T}^{(1)}_{02} &=& \frac{131 }{4} L_{\mu }-9 \zeta_3+\frac{2959}{24},
%----------------------------
\\
%----------------------------
{\cal T}^{(1)}_{03} &=& \frac{719}{20} L_{\mu }-9 \zeta_3+\frac{85339}{600},
%----------------------------
\\
%----------------------------
{\cal T}^{(1)}_{04} &=& \frac{769 }{20} L_{\mu }-9 \zeta_3+\frac{47537}{300},
%----------------------------
\\
%----------------------------
{\cal T}^{(1)}_{11} &=& \frac{145}{4}  L_\mu+\frac{523}{4},
%----------------------------
\\
%----------------------------
{\cal T}^{(1)}_{12} &=& \frac{163 }{4} L_{\mu } -\frac{351 \zeta_3}{35} +\frac{48231}{280},
%----------------------------
\\
%----------------------------
{\cal T}^{(1)}_{13} &=& \frac{879 }{20} L_{\mu }-27 \zeta_3+\frac{259573}{1200},
%----------------------------
\\
%----------------------------
{\cal T}^{(1)}_{14} &=& \frac{929 }{20} L_{\mu } -27 \zeta_3 +\frac{141499}{600},
%----------------------------
\\
%----------------------------
{\cal T}^{(1)}_{22} &=& \frac{181}{4} L_{\mu }+\frac{54 \zeta_3}{7}+\frac{20563}{112},
%----------------------------
\\
%----------------------------
{\cal T}^{(1)}_{23} &=& \frac{969 }{20}L_{\mu }-\frac{108 \zeta_3}{7}+\frac{997123}{4200},
%----------------------------
\\
%----------------------------
{\cal T}^{(1)}_{24} &=& \frac{1019 }{20}L_{\mu }-54 \zeta_3+\frac{732221}{2400},
%----------------------------
\\
%----------------------------
{\cal T}^{(1)}_{33} &=& \frac{1033 }{20} L_{\mu }+30 \zeta_3+\frac{2274761}{10800},
%----------------------------
\\
%----------------------------
{\cal T}^{(1)}_{34} &=& \frac{1083 }{20} L_{\mu }-\frac{180 \zeta_3}{11}+\frac{22888579}{79200},
%----------------------------
\\
%----------------------------
{\cal T}^{(1)}_{44} &=& \frac{1133}{20}  L_{\mu }+\frac{810 \zeta_3}{11}+\frac{674701}{3300}.
%----------------------------
\eqa
%----------------------------
\end{subequations}
%----------------------------

At two-loop level, the
lowest-lying short-distance coefficient matrix elements read
%--------------------------
\begin{subequations}
%----------------------------
\bqa
%----------------------------
{\cal T}^{(2)}_{00} &=& \frac{729 L_{\mu }^2}{16}-(8 \zeta_3+\frac{35 \pi ^2 }{6}-\frac{2961
  }{8}) L_{\mu }+205 \zeta_5-\frac{3 \pi ^4}{20}-\frac{651
   \zeta_3}{2}-\frac{275 \pi
   ^2}{24}+821 ,
%----------------------------
\\
%----------------------------
{\cal T}^{(2)}_{01} &=& \frac{10961 L_{\mu }^2}{144}+\left(-\frac{137 \zeta_3}{5}+\frac{704879}{1080}-\frac{145 \pi ^2}{18}\right) L_{\mu }-\frac{1577 \zeta_5}{2}-\frac{4 \pi ^2 \zeta_3}{5}-\frac{251 \pi ^4}{1350} \nn\\
&+& \frac{16009 \zeta_3}{100}-\frac{20857 \pi
   ^2}{1080}+\frac{30154739}{16200},
%----------------------------
\\
%----------------------------
{\cal T}^{(2)}_{02} &=& \frac{6943 L_{\mu }^2}{72}+\left(-61 \zeta_3+\frac{387371}{432}-\frac{80 \pi ^2}{9}\right) L_{\mu }+\frac{155 \zeta_5}{2}-2 \pi ^2 \zeta_3-\frac{67 \pi ^4}{270}-\frac{107491 \zeta_3}{180}\nn\\
&-&\frac{5921 \pi ^2}{216}+\frac{254500067}{103680},
%----------------------------
\\
%----------------------------
{\cal T}^{(2)}_{03} &=& \frac{202039 L_{\mu }^2}{1800}+\left(-\frac{321 \zeta_3}{5}+\frac{58020191}{54000}-\frac{436 \pi ^2}{45}\right) L_{\mu }+\frac{165 \zeta_5}{2}-2 \pi ^2 \zeta_3-\frac{343 \pi ^4}{1350}\nn\\
&-&\frac{12785497 \zeta_3}{18900}-\frac{28237 \pi
   ^2}{900}+\frac{3042907421}{1008000},
%----------------------------
\\
%----------------------------
{\cal T}^{(2)}_{04} &=&  \frac{451403 L_{\mu }^2}{3600}+\left(-\frac{667 \zeta_3}{10}+\frac{33226643}{27000}-\frac{917 \pi ^2}{90}\right) L_{\mu }+\frac{355 \zeta_5}{4}-2 \pi ^2 \zeta_3-\frac{1397 \pi ^4}{5400}\nonumber\\
&-& \frac{76336859 \zeta_3}{100800}-\frac{740213 \pi^2}{21600}+\frac{257785535129}{72576000},\nn\\
{\cal T}^{(2)}_{11} &=& \frac{16385 L_{\mu }^2}{144}+\left(-8 \zeta_3+\frac{220117}{216}-\frac{185 \pi ^2}{18}\right) L_{\mu }+3546 \zeta_5-\frac{3 \pi ^4}{20}-\frac{56177 \zeta_3}{20}-\frac{2711 \pi ^2}{108}+\frac{4892983}{2592},
%----------------------------
\\
%----------------------------
{\cal T}^{(2)}_{12} &=& \frac{9943 L_{\mu }^2}{72}+\left(-\frac{2659 \zeta_3}{35}+\frac{20948881}{15120}-\frac{100 \pi ^2}{9}\right) L_{\mu }-\frac{11037 \zeta_5}{2}-\frac{78 \pi ^2 \zeta_3}{35} -\frac{869 \pi ^4}{3150} \nn\\
&+& \frac{81199421 \zeta_3}{29400}-\frac{294449 \pi^2}{7560}+\frac{140650728527}{25401600},
%----------------------------
%----------------------------
\\
%----------------------------
{\cal T}^{(2)}_{13} &=&  \frac{31351 L_{\mu }^2}{200}+\left(-\frac{1003 \zeta_3}{5}+\frac{15990101}{9000}-\frac{536 \pi ^2}{45}\right) L_{\mu }+\frac{495 \zeta_5}{2}-6 \pi ^2 \zeta_3-\frac{38 \pi ^4}{75}\nn\\
&-& \frac{39593063 \zeta_3}{25200}-\frac{82603 \pi
   ^2}{1350}+\frac{208836835727}{36288000},
%----------------------------
\\
%----------------------------
{\cal T}^{(2)}_{14} &=& \frac{619643 L_{\mu }^2}{3600}+\left(-\frac{2081 \zeta_3}{10}+\frac{107582681}{54000}-\frac{1117 \pi ^2}{90}\right)
L_{\mu }+\frac{1035 \zeta_5}{4}-6 \pi ^2 \zeta_3-\frac{937 \pi ^4}{1800}\nn\\
&-& \frac{59475599 \zeta_3}{33600}-\frac{470201 \pi^2}{7200}+\frac{1450125210497}{217728000},
%----------------------------
\\
%----------------------------
{\cal T}^{(2)}_{22} &=& \frac{23711 L_{\mu }^2}{144}+\left(\frac{337 \zeta_3}{7}+\frac{9519367}{6048}-\frac{215 \pi ^2}{18}\right)
L_{\mu }+\frac{281325 \zeta_5}{14}+\frac{12 \pi ^2 \zeta_3}{7}-\frac{29 \pi ^4}{630}\nn\\
&-& \frac{20454703 \zeta_3}{1470}-\frac{37249 \pi
   ^2}{1512}-\frac{977775959}{5080320},
%----------------------------
\\
%----------------------------
{\cal T}^{(2)}_{23} &=& \frac{73967 L_{\mu }^2}{400}+\left(-\frac{4402 \zeta_3}{35}+\frac{28919281}{14000}-\frac{1147 \pi^2}{90}\right) L_{\mu }-\frac{44685 \zeta_5}{2}-\frac{24 \pi^2 \zeta_3}{7}-\frac{773 \pi ^4}{2100}\nn\\
&+& \frac{1162960399 \zeta_3}{88200}-\frac{1365911
   \pi ^2}{25200}+\frac{3573525754441}{285768000},
%----------------------------
\\
%----------------------------
{\cal T}^{(2)}_{24} &=&  \frac{90691 L_{\mu }^2}{450}+\left(-\frac{2176 \zeta_3}{5}+\frac{48201937}{18000}-\frac{596 \pi ^2}{45}\right) L_{\mu }+\frac{915 \zeta_5}{2}-12 \pi ^2 \zeta_3-\frac{847 \pi ^4}{900}\nn\\
&-&\frac{25432877 \zeta_3}{8400}-\frac{2216813 \pi
   ^2}{21600}+\frac{1065085696391}{108864000},
%----------------------------
\\
%----------------------------
{\cal T}^{(2)}_{33} &=& \frac{742727 L_{\mu }^2}{3600}+\left(\frac{695 \zeta_3}{3}+\frac{1897099387}{972000}-\frac{1219 \pi ^2}{90}\right) L_{\mu }+\frac{144475 \zeta_5}{2}+\frac{20 \pi ^2 \zeta_3}{3}+\frac{119 \pi ^4}{405}\nn\\
&-&\frac{91161661 \zeta_3}{1890}-\frac{74287 \pi
   ^2}{16200}-\frac{773409858259}{68040000},
%----------------------------
\\
%----------------------------
{\cal T}^{(2)}_{34} &=&  \frac{11191 L_{\mu }^2}{50}+\left(-\frac{1576 \zeta_3}{11}+\frac{1585362629}{594000}-\frac{632 \pi ^2}{45}\right) L_{\mu }-\frac{1465225 \zeta_5}{22}-\frac{40 \pi ^2 \zeta_3}{11}-\frac{793 \pi^4}{1980}\nn\\
&+& \frac{1689813617 \zeta_3}{40656}-\frac{44720293 \pi ^2}{712800}+\frac{2603110257632537}{98794080000},
%----------------------------
\\
%----------------------------
{\cal T}^{(2)}_{44} &=& \frac{871277 L_{\mu }^2}{3600}+\left(\frac{6833 \zeta_3}{11}+\frac{605198897}{297000}-\frac{1309 \pi ^2}{90}\right) L_{\mu }+\frac{4380975 \zeta_5}{22}+\frac{180 \pi ^2 \zeta_3}{11}+\frac{67 \pi^4}{66}\nn\\
&-& \frac{8897578969 \zeta_3}{67760}+\frac{511331 \pi ^2}{9900}-\frac{11028087335760793}{263450880000}.
%----------------------------
\eqa
%----------------------------
\end{subequations}
%----------------------------

%----------------------------------------

\end{document}